\documentclass[journal]{IEEEtran}

\usepackage{cite}
\usepackage{booktabs}
\usepackage{amssymb}
\usepackage{graphicx}
\usepackage{multirow}
\usepackage[table]{xcolor}
\usepackage{placeins}
\usepackage[hidelinks]{hyperref}
\ifCLASSINFOpdf
\else
\fi

\usepackage{amsmath}

\begin{document}

\title{Multi-View Speech Representation Learning for Parkinson’s Disease Detection Using Context-guided Cross-modal Attention}

\author{George Theodosiou$^\dagger$, Loukas Ilias$^\dagger$, Dimitris Askounis
\thanks{The authors are with the Decision Support Systems Laboratory, School of Electrical and Computer Engineering, National Technical University of Athens, 15780 Athens, Greece (e-mail: georgetheo19@gmail.com; lilias@epu.ntua.gr; askous@epu.ntua.gr). \\ $\dagger$The first two authors contributed equally.}}


\maketitle

\begin{abstract}
Parkinson’s disease (PD) is a progressive neurodegenerative disorder that frequently causes speech impairments associated with hypokinetic dysarthria. As speech production relies on the precise coordination of complex neuromuscular mechanisms, speech analysis has emerged as a promising non-invasive and cost-effective biomarker for early PD detection. Recent deep learning approaches have shown encouraging results; however, most existing methods rely on a single speech representation, potentially overlooking complementary pathological information encoded across different feature spaces. In this work, we propose a multi-branch deep learning framework for automatic Parkinson’s disease detection from speech. Each recording is segmented into 5-second chunks and represented using three complementary modalities: Log-Mel spectrograms, Mel-Frequency Cepstral Coefficients (MFCCs), and contextualized HuBERT embeddings extracted from raw waveforms. The spectrograms are processed using a pre-trained ResNet-18 encoder, MFCC sequences are modeled through a Bidirectional Long Short-Term Memory (BiLSTM) network, and raw speech is encoded using a pre-trained HuBERT foundation model. To effectively integrate these heterogeneous representations, we introduce a context-guided cross-modal attention mechanism that dynamically weights temporal HuBERT embeddings according to the global acoustic context derived from the spectrogram and MFCC branches. Experiments conducted on the publicly available Spanish PC-GITA corpus under strict speaker-independent 5-fold cross-validation demonstrate the effectiveness of the proposed approach. The proposed architecture achieves an accuracy of 91.51\%, an F1-score of 91.24\%, and an AUC of 95.97\%. Furthermore, ablation studies confirm the contribution of both the proposed context-guided cross-modal attention mechanism and the integration of complementary speech representations. These findings highlight the potential of heterogeneous speech modeling for robust and clinically reliable Parkinson’s disease detection.
\end{abstract}


\IEEEpeerreviewmaketitle

\section{Introduction}

Parkinson’s disease (PD) is a progressive neurodegenerative disorder characterized by the degeneration of dopaminergic neurons, leading to impaired motor control \cite{zafar2023parkinson,dauer2003mechanisms}. Among its early manifestations is hypokinetic dysarthria, a speech disorder associated with reduced vocal intensity, imprecise articulation, and impaired prosody \cite{atalar2023hypokinetic,moyagale2019dysarthria}. Because speech production requires the precise coordination of more than one hundred respiratory, laryngeal, and articulatory muscles, it is highly sensitive to neurological dysfunction \cite{simonyan2011laryngeal,tremblay2015neurobiology}. Consequently, subtle disease-related changes can often be detected through speech analysis, making speech a promising source of digital biomarkers for Parkinson’s disease. Furthermore, digital speech assessment offers a non-invasive, low-cost, and objective alternative to conventional clinical evaluations, facilitating large-scale screening and longitudinal disease monitoring \cite{rusz2024prodromal,robin2020speech}.

Existing research initiatives increasingly rely on automated speech analysis, as conventional clinical assessment is often subjective, time-consuming, and resource-intensive. Although recent advances in Deep Learning and self-supervised learning have enabled the extraction of complex, non-linear speech representations, several limitations remain. First, many existing studies rely on a single representation of the speech signal, such as spectrograms, handcrafted acoustic features, or foundation-model embeddings. However, speech is inherently multidimensional, containing complementary articulatory, phonatory, prosodic, and temporal information that may not be fully captured by any single representation. Second, when multiple feature streams are considered, they are often combined through simple concatenation or static fusion strategies that fail to explicitly model interactions between heterogeneous modalities and may overlook subtle pathological patterns. As a result, potentially valuable complementary information remains underutilized. These limitations motivate the development of multi-view learning frameworks capable of integrating diverse speech representations and dynamically identifying the most informative disease-related characteristics for robust and clinically reliable Parkinson’s disease detection.

To tackle the aforementioned limitations, we propose a novel multi-branch Deep Learning architecture for the automatic detection of Parkinson's disease. Instead of relying on a one-dimensional analysis of the audio, the proposed network simultaneously extracts and synthesizes information from three different streams: 1) Log-Mel spectrograms, 2) Mel-Frequency Cepstral Coefficients (MFCCs), and 3) contextualized speech embeddings obtained from the pre-trained HuBERT foundation model. For the optimal fusion of these heterogeneous features, we implement a context-guided cross-modal attention mechanism. This mechanism allows the network to dynamically focus on the most diagnostically informative temporal regions within the HuBERT representation using contextual information derived from the spectrogram and MFCC branches. We test our proposed approach on the Spanish PC-GITA dataset \cite{orozco-arroyave-etal-2014-new}. To ensure clinical reliability and prevent data leakage, all experiments are designed and executed using strict speaker-independent cross-validation protocols. We demonstrate that our multi-branch architecture achieves highly competitive performance, confirming its diagnostic capability. 

Our main contributions can be summarized as follows:
\begin{itemize}
    \item We introduce a novel multi-branch deep learning architecture that simultaneously processes three distinct streams of information: Spectrograms, MFCCs, and raw audio embeddings from HuBERT.
    \item We employ a context-guided cross-modal attention mechanism for the optimal fusion of heterogeneous speech data, which dynamically evaluates and focuses on the pathological speech patterns.
    \item We perform a series of ablation experiments to investigate the effectiveness of the proposed model.
\end{itemize}

\section{Related Work}

\subsection{Machine Learning-Based Methods}
Existing research initiatives for detecting Parkinson’s disease initially built on feature extraction and the training of shallow machine learning (ML) classifiers. Hireš et al. \cite{hires2023} investigated PD detection from speech using handcrafted acoustic features combined with XGBoost and evaluated the cross-dataset generalization capability of the proposed approach. To improve representation, Zebidi et al. \cite{zebidi2025} extracted MFCCs, STFT, and Mel-spectrograms, along with their derivatives, and converted them into 1D feature vectors; the authors employed the XGBoost classifier combined with the SMOTE technique for class balancing. Other researchers focused on specific physiological or articulatory markers. Pah et al. \cite{pah2022} analyzed vowels by extracting Vocal Tract Length (VTL) features and trained a support vector machine (SVM) classifier, demonstrating that VTL outperformed traditional glottal vibration parameters. Similarly, Narendra et al. \cite{narendra2021} extracted glottal features via iterative adaptive inverse filtering (IAIF) and quasi-closed phase (QCP) analysis, subsequently training an SVM.

Moving beyond vowels, studies have explored diadochokinetic (DDK) tasks and isolated words. For instance, the authors in \cite{hovsepyan2024} extracted syllable-level features from DDK tasks and trained an SVM, noting the specific articulatory importance of peak-based (consonant) segmentation. Oliveira et al. \cite{oliveira2025} trained a stacked ensemble model for disease severity estimation based on DisVoice features by uniquely dividing the multiclass severity problem into four separate binary classification tasks. Amato et al. \cite{amato2021} performed a multi-level analysis on isolated words—extracting features from the entire signal, voiced segments, and transition regions (onset/offset)—and utilized a k-nearest neighbor (KNN) classifier with an early fusion strategy.

Finally, several studies have explored the diagnostic value of connected speech. Favaro et al. \cite{favaro2023} conducted a statistical analysis of various interpretable biomarkers, analyzing features like the variability of the fundamental frequency during text reading across multiple languages. Furthermore, Scimeca et al. \cite{scimeca2023} demonstrated the benefit of incorporating connected speech alongside simpler tasks; the authors combined features from phonetically balanced sentences and sustained vowels using a KNN classifier, highlighting the importance of Mel-Frequency Cepstral Coefficients (MFCCs). Advanced signal processing techniques have also been applied to connected speech tasks; for example, Kiran Reddy et al. \cite{10.1121/10.0036660} applied a Wavelet Scattering Network to generate noise-robust coefficients, while Reddy and Alku \cite{reddy2023} introduced an exemplar-based sparse representation classification approach for sentence reading tasks, successfully bypassing the time-consuming hyperparameter tuning typically required by traditional ML models.

López et al. \cite{lopez19_interspeech} employed Fisher Vector representations derived from MFCC, articulatory, and prosodic features extracted from PC-GITA speech recordings. The study investigated both feature-level and task-level fusion across read text, monologue, DDK, and sentence tasks.

\subsection{Deep Learning Approaches}
The transition from handcrafted features to deep learning (DL) models enables networks to automatically learn complex representations directly from speech signals. A widely adopted approach involves converting speech into two-dimensional time-frequency representations and processing them with Convolutional Neural Networks (CNNs). Hireš et al. \cite{hires2023}, who also evaluated XGBoost, trained an Xception CNN on log-frequency power spectrograms extracted from sustained vowels; however, the model's performance degraded substantially when tested on unseen datasets, highlighting the generalization challenge. Bhatt et al. \cite{bhatt2023} replaced standard spectrograms with the High-resolution Superlet Transform, feeding the resulting representations into deep architectures such as InceptionResNetV2 and VGG-16. Janbakhshi and Kodrasi \cite{janbakhshi2021} advanced this by proposing a supervised CNN auto-encoder framework; by incorporating adversarial training with an auxiliary speaker identification task, they extracted representations that are simultaneously discriminative for the disease and invariant to the speaker's identity.

Given the limited size of PD speech corpora, several studies have turned to transfer learning to leverage knowledge from pre-trained models. Zahid et al. \cite{zahid2020} fine-tuned the pre-trained AlexNet network to extract deep features from spectrogram images across multiple speech tasks, outperforming classical handcrafted methods. Ibarra et al. \cite{ibarra2023} adopted a domain adversarial training strategy, adapting architectures like Time-CNN-LSTM originally trained on a different voice disorders database to multi-corpus scenarios to mitigate corpus-dependent clustering. Vásquez-Correa et al. \cite{vasquez2021} investigated transfer learning across languages and speech disorders by training CNN models on a source corpus and subsequently fine-tuning them on a target dataset. Using Mel-spectrogram representations extracted from speech onset transitions, the authors transferred learned CNN parameters across Spanish, German, and Czech speech corpora, as well as across different speech disorders.

Another line of research bypasses spectrogram computation entirely by processing raw waveforms in an end-to-end fashion. Narendra et al. \cite{narendra2021} proposed an end-to-end system combining a CNN with a multilayer perceptron (MLP) trained directly on glottal flow signals. Finally, Rey-Paredes et al. \cite{rey2025} compared time-series classification networks applied directly to raw waveforms and incorporated Generative Adversarial Networks (GANs), specifically the BigVSAN vocoder, to synthesize highly realistic voice samples for data augmentation, effectively mitigating the small dataset size.

Perrone et al. \cite{PERRONE2026108954} investigated Parkinson's disease voice classification using transformer-based deep learning models together with speech-related error metrics, including Word Error Rate (WER) and Character Error Rate (CER). Their study analyzed these metrics for both binary classification between Parkinsonian and healthy speakers and the assessment of disease severity.

Zeng et al. \cite{10.3389/fphys.2026.1806415} proposed a hybrid CNN-Mamba framework for Parkinson's disease severity classification using multimodal speech biomarkers transformed into two-dimensional representations. The model was trained and validated on speaker-disjoint Spanish PC-GITA data and further evaluated on independent Mandarin and public Parkinsonian speech corpora, with speaker-level decisions obtained by aggregating segment-level predictions.

\subsection{Self-Supervised Learning and Foundation Models}
The advent of self-supervised learning (SSL) and large-scale pre-trained foundation models has introduced a fundamentally different paradigm for PD detection from speech. Rather than relying on handcrafted acoustic measures, these models learn rich acoustic representations from vast amounts of unlabeled speech data, effectively addressing the small dataset challenge inherent to clinical PD corpora. Favaro et al. \cite{favaro2023dnn}, who also employed XGBoost as a classical baseline in the previous section, conducted an extensive comparison between handcrafted features and SSL-based representations. Their results demonstrated the superiority of the SSL approach across both mono-lingual and cross-lingual settings, with HuBERT embeddings achieving an AUC of 0.90 on PC-GITA. Klempir et al. \cite{klempir2025} investigated the evolution of the Wav2Vec architecture by comparing versions 1.0 and 2.0 across different speech tasks. Using a multi-criteria ranking approach, their analysis confirmed that the contextual representations produced by Wav2Vec 2.0's transformer layers lead to significantly higher diagnostic accuracy for connected speech, while its feature extractor layer proved optimal for vowels.

Despite their strong performance, SSL models often function as black boxes, limiting their clinical interpretability. To bridge this gap, Gimeno-Gómez et al. \cite{gimeno2025} integrated Wav2Vec 2.0 embeddings into an Interpretable Cross-Attention Framework, successfully combining the accuracy of SSL representations with the identification of specific clinically meaningful biomarkers at both the embedding and temporal levels. Following a similar motivation for finer-grained analysis, Gallo-Aristizábal et al. \cite{gallo2024} performed fine-tuning of Wav2Vec 2.0 at sub-lexical levels. Their study revealed that analysis at the phoneme and syllable level substantially outperforms word-level analysis, identifying nasals and plosives as the most discriminative phonological classes affected by the disease.

As the number of available foundation models continues to grow, broader comparative evaluations have been conducted. La Quatra et al. \cite{laquatra2024} systematically evaluated widely used models, identifying WavLM Base as the best-performing model on the PC-GITA corpus, particularly when evaluated under real-world operative conditions employing speech enhancement techniques. Exploring data diversity, Zhong et al. \cite{zhong2025} demonstrated that models fed with WavLM Base representations through an attention pooling mechanism can achieve promising classification results even when trained on non-diagnostic speech data compared to purely diagnostic corpora like PC-GITA. 

La Quatra et al. \cite{10889445} proposed a bilingual dual-head architecture for Parkinson's disease detection from speech, using separate task-specific heads for diadochokinetic and continuous speech tasks. Their model combines SSL-based speech representations with wavelet-derived features and incorporates adaptive layers, convolutional bottlenecks, and contrastive learning to improve cross-language generalization between Slovak and Spanish speech corpora. Purohit et al. The authors in \cite{10890852} investigated speech foundation models for Parkinson's disease detection on the PC-GITA corpus by comparing frozen feature extraction, full fine-tuning, and parameter-efficient fine-tuning. The authors introduced a cross-validation-based layer selection strategy and evaluated Wav2Vec2 Base, XLSR, and Whisper-small representations, including LoRA-based adaptation for efficient model tuning.

\subsection{Related Work Review Findings}
Existing research initiatives on automatic PD detection from speech have followed a clear methodological trajectory. Early approaches relied on the extraction of handcrafted acoustic features, such as jitter, shimmer, MFCCs, and fundamental frequency statistics, combined with shallow classifiers including SVMs, KNN, and XGBoost. While these methods offer a degree of clinical interpretability, they require substantial domain expertise for feature engineering and often exhibit limited generalization across different datasets and languages. More recent approaches employ deep learning architectures operating on spectrogram representations, transfer learning from pre-trained vision models, or end-to-end frameworks that process raw waveforms directly. These methods reduce the need for manual feature design and enable the automatic extraction of complex acoustic patterns, but they remain largely constrained to a single representation of the speech signal. The latest generation of approaches leverages SSL foundation models, such as Wav2Vec 2.0, HuBERT, and WavLM, which learn rich contextualized speech representations from large quantities of unlabeled data and have demonstrated promising performance and cross-lingual robustness.

Despite this progress, a critical limitation persists across much of the existing literature: most studies rely on a single representation of the speech signal. Models trained exclusively on spectrograms effectively capture energy distribution patterns but may overlook fine articulatory dynamics encoded in cepstral representations. Likewise, approaches based solely on SSL embeddings, while highly effective, do not explicitly exploit the complementary information provided by traditional acoustic features. Consequently, the joint utilization of heterogeneous speech representations remains relatively underexplored compared to single-representation approaches. Furthermore, the optimal integration of such representations remains an open challenge, as simple fusion strategies such as feature concatenation may fail to capture complex interactions and diagnostically relevant temporal dependencies across feature streams. These observations motivate the development of architectures capable of simultaneously exploiting complementary speech representations and dynamically focusing on the most informative pathological patterns associated with Parkinson’s disease.

Therefore, our work differs from existing research initiatives in the following ways: 1) we propose a novel multi-branch deep learning architecture that simultaneously processes three complementary representations of the same speech signal, namely Log-Mel spectrograms, MFCC sequences, and HuBERT embeddings; 2) we introduce a context-guided cross-modal attention mechanism for the optimal fusion of these heterogeneous streams, enabling the network to dynamically focus on the most pathologically informative temporal regions of speech; and 3) we evaluate our proposed system under strict speaker-independent cross-validation conditions on the PC-GITA corpus, ensuring clinically reliable results that are not inflated by speaker identity leakage.

\section{Methodology}

\begin{figure*}
    \centering
    \includegraphics[width=0.66\linewidth]{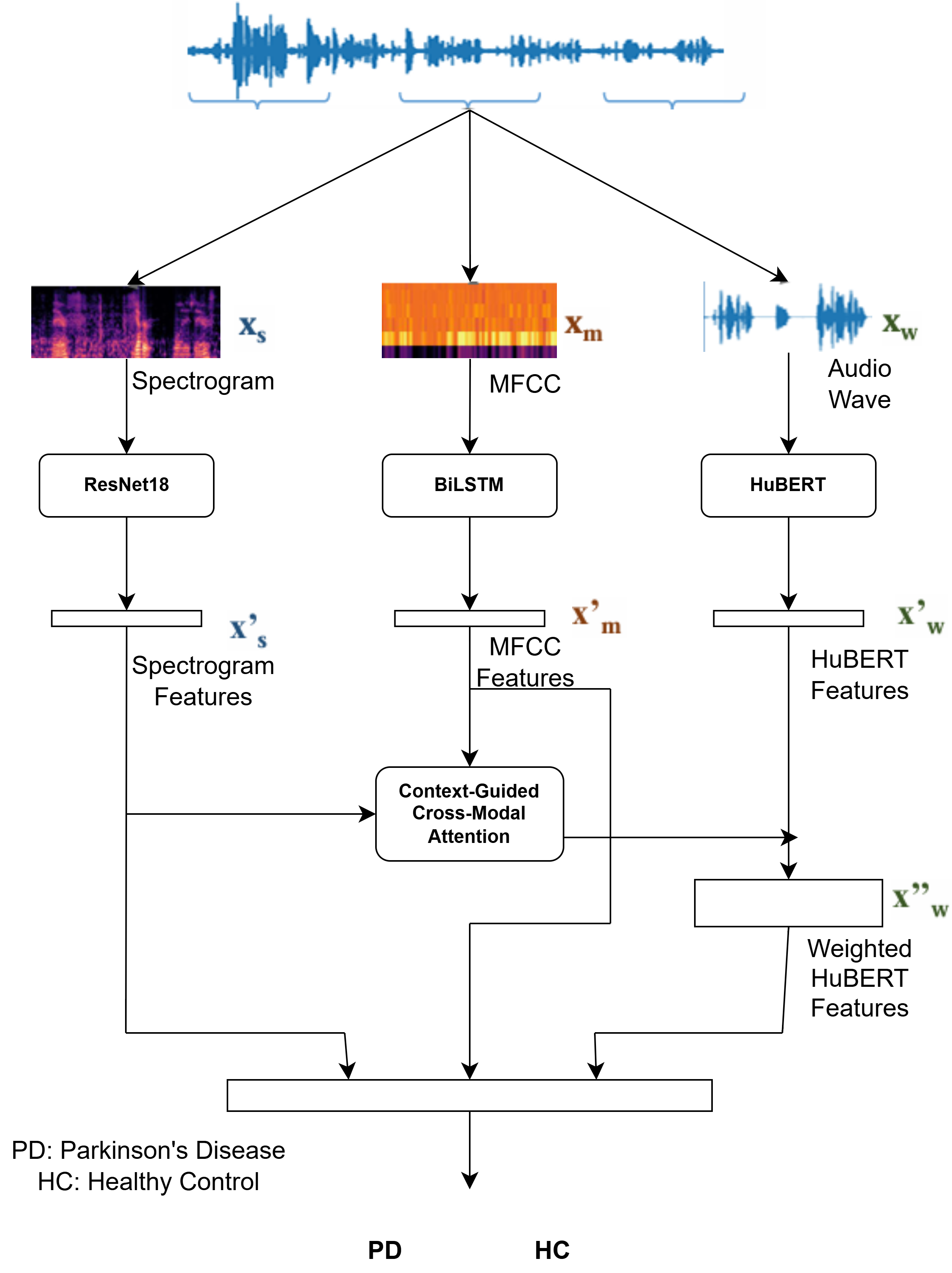}
    \caption{Proposed Methodology}
    \label{fig:architecture}
\end{figure*}

In this section, we present our proposed framework for automatic PD detection from speech signals. Motivated by the observation that pathological speech characteristics manifest across multiple acoustic levels, we adopt a multi-view representation learning strategy that simultaneously exploits complementary information derived from spectral, cepstral, and self-supervised speech representations. The overall architecture is illustrated in Fig.~\ref{fig:architecture}.

Unlike conventional approaches that rely on a single representation of the speech signal, our framework processes three complementary modalities extracted from each audio segment: (i) Log-Mel spectrograms, which preserve detailed time-frequency structures, (ii) Mel-Frequency Cepstral Coefficients (MFCCs), which capture vocal tract and articulatory characteristics, and (iii) contextualized HuBERT embeddings obtained directly from the raw waveform. These heterogeneous representations are subsequently integrated through a context-guided attention mechanism that enables the model to focus on diagnostically informative temporal regions of speech.

\subsection{Input Representations}

Each recording is first segmented into fixed-length 5-second chunks. All speech signals are resampled to 16 kHz, resulting in segments of 80,000 samples.

For every chunk, three complementary representations are extracted.

\textbf{1) Log-Mel Spectrogram:}
A Short-Time Fourier Transform (STFT) is applied using a 40 ms Hamming window and a 10 ms hop size. The resulting power spectrum is projected onto 200 Mel-frequency filter banks and converted to the logarithmic domain. To capture temporal dynamics, the Log-Mel representation is augmented with first-order and second-order temporal derivatives (delta and delta-delta coefficients). The three channels are stacked to form an RGB-like representation:

\begin{equation}
x_s \in \mathbb{R}^{3 \times 224 \times 224}
\end{equation}

after interpolation to \(224 \times 224\), Min-Max normalization, and ImageNet standardization.

\textbf{2) MFCC Sequence:}
The first 40 Mel-Frequency Cepstral Coefficients are extracted using the same STFT parameters. MFCCs capture the spectral envelope of speech and provide a compact representation of vocal tract characteristics. Each chunk is represented as

\begin{equation}
x_m \in \mathbb{R}^{T_m \times 40}
\end{equation}

where \(T_m\) denotes the number of temporal frames.

\textbf{3) Raw Audio Waveform:}
The original waveform is retained as

\begin{equation}
x_w \in \mathbb{R}^{L}, \quad L = 80,000
\end{equation}

corresponding to a 5-second segment sampled at 16 kHz. A binary attention mask is generated in parallel to identify valid speech samples and suppress padded regions during processing.

\subsection{Modality-Specific Encoders}

Each representation is processed independently by a dedicated encoder.

\textbf{Spectrogram Branch:}
The spectrogram representation is fed to a ResNet-18 network pre-trained on ImageNet. The final classification layer is removed and the network is used purely as a feature extractor:

\begin{equation}
x'_s = \mathrm{ResNet18}(x_s)
\in \mathbb{R}^{512}
\end{equation}

This branch captures high-level spectral and time-frequency patterns associated with Parkinsonian speech.

\textbf{MFCC Branch:}
The MFCC sequence is processed by a Bidirectional Long Short-Term Memory (BiLSTM) network with 64 hidden units per direction. The final forward and backward hidden states are concatenated and projected through a linear layer:

\begin{equation}
x'_m = \mathrm{BiLSTM}(x_m)
\in \mathbb{R}^{128}
\end{equation}

Dropout is applied before and after the projection layer to improve generalization.

\textbf{HuBERT Branch:}
The raw waveform is processed using the pre-trained HuBERT-Base-ls960 foundation model. HuBERT is kept frozen during training and serves exclusively as a feature extractor. The model produces a sequence of contextualized frame-level speech representations:

\begin{equation}
x'_w = \mathrm{HuBERT}(x_w)
\in \mathbb{R}^{T_w \times 768}
\end{equation}

where \(T_w\) denotes the number of HuBERT frames.

Unlike the spectrogram and MFCC branches, which are compressed into global representations, the HuBERT branch preserves temporal resolution throughout the fusion stage.

\subsection{Context-Guided Cross-Modal Attention}

The outputs of the spectrogram and MFCC branches are concatenated to form a global acoustic context vector:

\begin{equation}
x_{co}
=
[x'_s;x'_m]
\in
\mathbb{R}^{640}
\end{equation}

This representation summarizes complementary spectral and cepstral information extracted from the speech segment.

Rather than treating all HuBERT frames equally, the global acoustic context is used to identify the most diagnostically informative temporal regions within the HuBERT sequence. The global context is projected into a latent query representation:

\begin{equation}
q = W_q x_{co}
\end{equation}

while each HuBERT frame embedding is projected into the same latent space:

\begin{equation}
k_t = W_k x'_{w,t}
\end{equation}

Attention scores are computed through dot-product similarity:

\begin{equation}
e_t = q^\top k_t
\end{equation}

and normalized using a softmax operation:

\begin{equation}
\alpha_t
=
\frac{\exp(e_t)}
{\sum_j \exp(e_j)}
\end{equation}

Padding positions are masked prior to normalization.

The final attended HuBERT representation is obtained as a weighted aggregation of all HuBERT frames:

\begin{equation}
x''_w
=
\sum_t \alpha_t x'_{w,t}
\in
\mathbb{R}^{768}
\end{equation}

Through this mechanism, the network dynamically focuses on the HuBERT frames that are most consistent with the global acoustic characteristics captured by the spectrogram and MFCC branches.

\subsection{Multimodal Fusion and Classification}

The attended HuBERT representation is concatenated with the global acoustic context:

\begin{equation}
x
=
[x_{co};x''_w]
\in
\mathbb{R}^{1408}
\end{equation}

The fused feature vector is subsequently processed by a Multi-Layer Perceptron (MLP) consisting of a fully connected layer with 64 hidden units, a ReLU activation function, a dropout layer with probability 0.2, and a final classification layer producing two output logits corresponding to the Parkinson’s Disease (PD) and Healthy Control (HC) classes.

The final posterior probabilities are obtained through softmax normalization, and the network is optimized using the standard Cross-Entropy loss function computed at the chunk level.

\subsection{Subject-Level Decision Aggregation}

Although optimization is performed at the chunk level, the clinical objective is subject-level diagnosis. During inference, every 5-second chunk belonging to the same subject is independently processed by the network, producing a Parkinson’s disease probability \(P_i\).

The final subject-level probability is obtained through Mean Probability Aggregation (MPA):

\begin{equation}
P_{\mathrm{subject}}
=
\frac{1}{N}
\sum_{i=1}^{N}
P_i
\end{equation}

where \(N\) denotes the number of chunks associated with the subject.

A subject is classified as Parkinson’s Disease if

\begin{equation}
P_{\mathrm{subject}} > 0.5
\end{equation}

and as Healthy Control otherwise.

This aggregation strategy reduces the influence of noisy or weakly informative chunks and produces a more robust subject-level diagnosis.

\section{Experiments}

\subsection{Dataset}
The PC-GITA dataset contains speech recordings in Spanish from 50 Parkinson's disease (PD) patients (25 male and 25 female) and 50 healthy control (HC) speakers (25 male and 25 female). The data was sampled at 44.1 kHz with a resolution of 16 bits. The PD patients were diagnosed by neurologists and assessed using the MDS-UPDRS-III scale, while the healthy controls are free of any reported PD symptoms or other neurodegenerative conditions. Speaker age ranges from 31 to 86 years old. The recordings were performed in a sound-proof booth, ensuring controlled acoustic conditions. The database includes speech collected using the following tasks: (1) sustained phonation of vowels, (2) reading 25 isolated words aloud, (3) diadochokinetic (DDK) exercises involving the repetition of syllable sequences such as /pa-ta-ka/ and /pe-ta-ka/, (4) reading sentences aloud, (5) reading a text consisting of a dialogue between a doctor and a patient, and (6) giving a spontaneous monologue about daily activities. 

In this study, we use the text reading and monologue tasks, as connected speech has been shown to carry richer diagnostic information compared to isolated phonation tasks. All speech signals are downsampled to 16 kHz prior to feature extraction.

\subsection{Experimental Setup}
We use the Adam optimizer with a learning rate of $10^{-4}$. We use a batch size of 64. The network is optimized using the standard Cross-Entropy loss function, which is computed at the 5-second chunk level during the training phase. We employ a Stratified Group 5-Fold Cross-Validation strategy in order to prevent data leakage. In each outer fold, approximately 80\% of subjects are used for training and 20\% for testing. Within the training fold, a further speaker-level split reserves 15\% of training subjects as an internal early stopping set. The models were trained for a maximum of 15 epochs. Training halts automatically if the F1-score on the early stopping set does not improve for 3 consecutive epochs, after which the best-performing model weights are restored and evaluated on the test set. During the testing phase, since the clinical objective is subject-level diagnosis, the chunk-level softmax probabilities are aggregated using Mean Probability Aggregation to produce a final prediction for each speaker. We use the PyTorch framework for implementing and training our deep learning models. Additionally, we utilize the Hugging Face \texttt{transformers} library. All experiments were accelerated using a single NVIDIA A100 PCIe 80 GB GPU.

\subsection{Evaluation Metrics}
Precision, Recall, F1-score, Accuracy, Specificity, and AUROC are reported. Because our evaluation relies on a 5-fold cross-validation scheme, we report the average values and standard deviations for each of these metrics across all test folds.

\subsection{Baselines}
We compare the proposed multi-branch context-guided cross-modal attention architecture against representative published methods evaluated on the PC-GITA corpus. The selected baselines are relevant because they focus on connected-speech or task-specific PC-GITA recordings, including text reading and monologue tasks, rather than relying exclusively on sustained vowels or DDK phonation. 

\begin{enumerate} 

\item \textbf{Foundation-model baseline:} La Quatra et al. \cite{10889445} proposed a bilingual dual-head deep architecture for Parkinson's disease detection from speech. This method uses self-supervised speech representations together with wavelet-derived features and employs task-specific heads for DDK and continuous-speech recordings. 

\item \textbf{Parameter-efficient fine-tuning baseline:} Purohit et al. \cite{10890852} investigated the adaptation of speech foundation models for Parkinson's disease detection on PC-GITA. This method compares layer selection, full fine-tuning, and parameter-efficient fine-tuning using LoRA, with the best configuration based on Whisper fine-tuned with LoRA of rank 16. 

\item \textbf{Text-reading baseline:} The method in \cite{10.1121/10.0036660} uses the PC-GITA text-reading task for Parkinson's disease detection. This approach extracts speech representations from reading recordings and trains a feed-forward neural network (FFNN) classifier. 

\item \textbf{Monologue baseline:} López et al. \cite{lopez19_interspeech} focused on the monologue task from PC-GITA. This method uses MFCC-based and articulation-related acoustic features to characterize spontaneous speech patterns associated with Parkinson's disease.

\end{enumerate}

\textit{We note that direct numerical comparisons should be interpreted with caution, as different studies may adopt varying cross-validation protocols and subject-independence constraints.}

\section{Results}

The subject-level classification results are presented in Table~\ref{tab:results}. The proposed multi-branch context-guided cross-modal attention framework achieved the best overall performance among the considered baselines, obtaining an accuracy of (91.51\%), an F1-score of (91.24\%), a precision of (93.99\%), a specificity of (93.89\%), and an AUROC of (95.97\%). The relatively low standard deviations across folds indicate stable performance under the speaker-independent evaluation protocol.

\begin{table}[!htb]
\centering
\caption{Subject-level classification results.
Mean $\pm$ standard deviation across folds are reported. P: Precision, R: Recall, F1: F1-score, A: Accuracy, S: Specificity, AUC: AUROC}
\label{tab:results}
\begin{tabular}{lcccccc}
\toprule
\textbf{Model} & \textbf{P.} & \textbf{R.}
& \textbf{F1} & \textbf{A.}
& \textbf{S.} & \textbf{AUC} \\
\midrule
\cite{10889445} 
&  & 91.67
& 90.00 & 90.00
& 88.33 &\\ \hline

\cite{10890852}  &   & 86.36 & 85.34  & 85.00   & 83.63  & \\

&   & $\pm$ 10.1 & $\pm$ 8.90 & $\pm$ 9.60  & $\pm$ 15.20 & \\ \hline

\cite{10.1121/10.0036660} &  &   &    &  87.00 & & 95.00\\ \hline

\cite{lopez19_interspeech}  &   &   &  84.00  & 84.00  & & 90.08 \\ 

\midrule
\textbf{Ours} & \textbf{93.99} & \textbf{89.00} & \textbf{91.24} & \textbf{91.51} & \textbf{93.89} & \textbf{95.97} \\
                & $\pm$3.62 & $\pm$4.90 & $\pm$1.51 & $\pm$1.14 & $\pm$3.86 & $\pm$2.32 \\
\bottomrule
\end{tabular}
\end{table}

Compared to recent foundation-model-based approaches, the proposed method consistently achieved higher performance. In particular, it outperformed the bilingual dual-head architecture of La Quatra et al.~\cite{10889445}, improving the F1-score from (90.00\%) to (91.24\%) and the accuracy from (90.00\%) to (91.51\%). Similarly, our framework surpassed the Whisper-LoRA model of Purohit et al.~\cite{10890852}, yielding improvements of approximately (6\%) in F1-score and (6.5\%) in accuracy.

The proposed method also compared favorably against task-specific approaches. The feed-forward neural network model of \cite{10.1121/10.0036660}, evaluated on the text-reading task, achieved an AUROC of (95.00\%), while our method reached (95.97\%). Likewise, the monologue-based approach of López et al.~\cite{lopez19_interspeech} reported an F1-score and accuracy of (84.00\%), substantially lower than those obtained by the proposed architecture.

Overall, these results suggest that combining complementary spectral, cepstral, and self-supervised speech representations within a unified attention-guided fusion framework provides a more discriminative characterization of Parkinsonian speech than approaches relying on a single feature representation or a single foundation model. The strong performance across all evaluation metrics further supports the effectiveness of the proposed multimodal design for subject-level Parkinson's disease detection from connected speech.

\section{Ablation Study}

\subsection{SSL Speech Models}

To investigate the impact of the self-supervised learning (SSL) backbone employed in the proposed architecture, we conducted an ablation study comparing two alternative speech foundation models, namely Wav2Vec 2.0 Base and Wav2Vec 2.0 XLSR-53. In each experiment, the SSL encoder branch of the proposed framework was replaced by the corresponding model, while all remaining components, including the spectrogram branch, MFCC branch, context-guided cross-modal attention module, training procedure, and evaluation protocol, were kept unchanged.

The results are presented in Table~\ref{ablation_ssl}. Both SSL backbones achieved strong performance, confirming the effectiveness of self-supervised speech representations for Parkinson’s disease detection. Wav2Vec 2.0 Base obtained an accuracy of 87.97\% and an F1-score of 87.37\%, while achieving a high specificity of 92.94\%, indicating a strong capability to correctly identify healthy control subjects. Replacing the encoder with Wav2Vec 2.0 XLSR-53 improved recall from 83.00\% to 89.00\%, leading to an increase in F1-score (88.97\%) and overall accuracy (89.01\%). This improvement may be attributed to the multilingual pre-training strategy of XLSR-53, which enables the model to learn more robust and diverse acoustic representations.

Despite these gains, the proposed HuBERT-based methodology consistently achieves the best overall performance. Specifically, it attains the highest precision (93.99\%), F1-score (91.24\%), accuracy (91.51\%), specificity (93.89\%), and AUC (95.97\%). While HuBERT and XLSR-53 achieve identical recall values (89.00\%), the proposed configuration provides a substantially better balance between sensitivity and precision, resulting in fewer false-positive predictions and superior overall classification performance.

These findings suggest that HuBERT constitutes the most suitable SSL backbone for the proposed multi-branch architecture. The masked prediction objective employed during HuBERT pre-training may enable the extraction of richer contextual and phonetic information, which can be effectively exploited by the context-guided cross-modal attention module when combined with the complementary spectral and cepstral representations. Furthermore, the lower standard deviations observed across folds indicate that the proposed HuBERT-based configuration produces more stable and reliable predictions under strict speaker-independent evaluation conditions.

\begin{table}[h!]
\caption{ABLATION STUDY (SSL SPEECH). BEST RESULTS PER EVALUATION METRIC ARE IN BOLD.}
\label{ablation_ssl}
\begin{center}
\resizebox{\columnwidth}{!}{%
\begin{tabular}{lcccccc}
\toprule
& \multicolumn{6}{c}{\textbf{Evaluation metrics}} \\
\cline{2-7}
\textbf{Architecture} & \textbf{Precision} & \textbf{Recall} & \textbf{F1-score} & \textbf{Accuracy} & \textbf{Specificity} & \textbf{AUC} \\
\midrule
\rowcolor{gray!30} \multicolumn{7}{l}{\textbf{Ablation Experiments}} \\
wav2vec2 Base & 92.63 & 83.00 & 87.37 & 87.97 & 92.94 & 95.89 \\
                & $\pm$4.93 & $\pm$4.00 & $\pm$1.88 & $\pm$2.05 & $\pm$5.19 & $\pm$2.55 \\
\hline
wav2vec2 xlsr-53& 89.47 & \textbf{89.00} & 88.97 & 89.01 & 89.03 & 95.87 \\
                & $\pm$6.10 & $\pm$6.63 & $\pm$4.06 & $\pm$4.02 & $\pm$6.49 & $\pm$2.20 \\
\midrule
\rowcolor{gray!30} \multicolumn{7}{l}{\textbf{Proposed Methodology}} \\
\textbf{Methodology} & \textbf{93.99} & \textbf{89.00} & \textbf{91.24} & \textbf{91.51} & \textbf{93.89} & \textbf{95.97} \\
                & $\pm$3.62 & $\pm$4.90 & $\pm$1.51 & $\pm$1.14 & $\pm$3.86 & $\pm$2.32 \\
\bottomrule
\end{tabular}%
}
\end{center}
\end{table}

\subsection{Fusion Methods}

To evaluate the effectiveness of the proposed context-guided cross-modal attention mechanism, we compare it against three representative multimodal fusion strategies. Unlike these approaches, which perform static feature fusion, the proposed methodology dynamically weights the temporal SSL representations using global acoustic context derived from the spectrogram and MFCC branches.

\begin{itemize}
\item \textbf{Concatenation}: The feature vectors extracted from the three branches are directly concatenated into a single representation and subsequently passed to the classification network. This serves as a simple baseline that does not explicitly model interactions between modalities.

\item \textbf{Multimodal Low-rank Bilinear Pooling (MLB)}: MLB is a bilinear fusion method that captures pairwise multiplicative interactions between multimodal representations through low-rank projections. By projecting each modality into a shared latent space before performing element-wise multiplication, MLB approximates full bilinear pooling while significantly reducing the number of trainable parameters and computational complexity.

\item \textbf{Multimodal Factorized Bilinear (MFB) pooling \cite{8334194}}: MFB is a bilinear fusion approach that captures multiplicative interactions between multimodal representations through factorized pooling operations. It aims to improve multimodal feature expressiveness while maintaining computational efficiency.

\end{itemize}

The results of the fusion ablation study are reported in Table~\ref{ablation_fusion}. Overall, all fusion strategies achieve competitive performance, demonstrating the benefit of combining heterogeneous speech representations. However, clear differences emerge in their ability to balance sensitivity and specificity.

The simple concatenation baseline achieves an accuracy of 88.51\% and an F1-score of 88.07\%, indicating that the three modalities contain complementary information that can be effectively exploited even without explicitly modeling cross-modal interactions. Interestingly, concatenation achieves the highest AUC (96.02\%), suggesting strong ranking capability across different decision thresholds. Nevertheless, its lower recall (85.00\%) indicates a reduced ability to correctly identify Parkinson’s disease patients compared to the proposed approach.

MLB achieves the highest precision (95.71\%) and specificity (95.94\%), demonstrating excellent discrimination of healthy control subjects. However, this improvement comes at the expense of recall, which decreases to 83.00\%. Consequently, although MLB achieves a slightly higher F1-score than simple concatenation, its overall performance remains below that of the proposed methodology. This behavior suggests that MLB produces a more conservative classifier that prioritizes minimizing false positives while missing a larger number of Parkinson’s disease cases.

MFB pooling exhibits the weakest overall performance among the evaluated fusion methods, achieving an accuracy of 87.51\% and an F1-score of 87.69\%. Furthermore, it presents the largest standard deviations across folds, particularly for precision and specificity, indicating reduced robustness and greater sensitivity to the specific speaker partition used during evaluation.

In contrast, the proposed context-guided cross-modal attention mechanism consistently achieves the strongest overall results, obtaining the highest recall (89.00\%), F1-score (91.24\%), and accuracy (91.51\%), while maintaining high precision (93.99\%) and specificity (93.89\%). Although the AUC obtained by the proposed methodology (95.97\%) is marginally lower than that achieved by simple concatenation (96.02\%), the difference is negligible and falls well within the observed standard deviations. More importantly, the proposed architecture provides a substantially better balance between sensitivity and specificity, resulting in superior classification performance at the operating point used for diagnosis.

These findings suggest that the performance gains achieved by the proposed methodology are not solely attributable to the extracted features, but also to the fusion strategy itself. By dynamically focusing on the most informative temporal regions of the SSL representations according to the global acoustic context, the context-guided cross-modal attention mechanism is able to exploit complementary information across modalities more effectively than static fusion approaches. Furthermore, the relatively low standard deviations observed across folds indicate that the proposed fusion strategy produces stable and reliable predictions under strict speaker-independent evaluation conditions.

\begin{table}[h!]
\caption{ABLATION STUDY (FUSION). BEST RESULTS PER EVALUATION METRIC ARE IN BOLD.}
\label{ablation_fusion}
\begin{center}
\resizebox{\columnwidth}{!}{%
\begin{tabular}{lcccccc}
\toprule
& \multicolumn{6}{c}{\textbf{Evaluation metrics}} \\
\cline{2-7}
\textbf{Architecture} & \textbf{Precision} & \textbf{Recall} & \textbf{F1-score} & \textbf{Accuracy} & \textbf{Specificity} & \textbf{AUC} \\
\midrule
\rowcolor{gray!30} \multicolumn{7}{l}{\textbf{Ablation Experiments}} \\
Concatenation   & 91.45 & 85.00 & 88.07 & 88.51 & 91.99 & \textbf{96.02} \\
                & $\pm$2.47 & $\pm$3.16 & $\pm$2.16 & $\pm$1.94 & $\pm$2.48 & $\pm$1.45 \\
\hline
MLB             & \textbf{95.71} & 83.00 & 88.75 & 89.50 & \textbf{95.94} & 95.86 \\
                & $\pm$3.84 & $\pm$4.00 & $\pm$1.20 & $\pm$1.01 & $\pm$3.90 & $\pm$1.75 \\
\hline
MFB             & 88.06 & 88.00 & 87.69 & 87.51 & 86.93 & 94.34 \\
                & $\pm$7.98 & $\pm$4.00 & $\pm$2.98 & $\pm$3.49 & $\pm$9.42 & $\pm$2.53 \\
\midrule
\rowcolor{gray!30} \multicolumn{7}{l}{\textbf{Proposed Methodology}} \\
\textbf{Methodology} & 93.99 & \textbf{89.00} & \textbf{91.24} & \textbf{91.51} & 93.89 & 95.97 \\
                & $\pm$3.62 & $\pm$4.90 & $\pm$1.51 & $\pm$1.14 & $\pm$3.86 & $\pm$2.32 \\
\bottomrule
\end{tabular}%
}
\end{center}
\end{table}

\subsection{Input Modalities}

To quantify the contribution of each input modality, we conducted an ablation study in which one of the acoustic branches was removed while keeping the remaining architecture, training procedure, and context-guided cross-modal attention mechanism unchanged. Specifically, we evaluated two reduced configurations: (i) Spectrogram + HuBERT Base and (ii) MFCC + HuBERT Base. These variants were subsequently compared against the complete three-branch architecture incorporating spectrograms, MFCCs, and HuBERT representations.

The results are presented in Table~\ref{ablation_modalities}. Both reduced configurations achieve competitive performance, confirming that each acoustic representation individually provides complementary information to the self-supervised HuBERT embeddings. However, a consistent performance degradation is observed whenever one of the acoustic branches is removed.

The Spectrogram + HuBERT configuration achieves an accuracy of 87.45\% and an F1-score of 86.62\%, outperforming the MFCC + HuBERT variant across all evaluation metrics. This observation suggests that the spectrogram branch contributes richer complementary information to the SSL representations than MFCCs alone. Spectrograms preserve detailed time-frequency structures, harmonic content, and energy distribution patterns that may capture subtle manifestations of hypokinetic dysarthria, including reduced articulatory precision and altered prosodic characteristics. Consequently, when combined with the contextual representations learned by HuBERT, spectrograms provide a stronger basis for disease discrimination.

The MFCC + HuBERT configuration achieves an accuracy of 85.98\% and an F1-score of 85.16\%. Although MFCCs effectively characterize the spectral envelope and vocal tract configuration, they represent a more compact and compressed description of the speech signal. As a result, some fine-grained acoustic details that may be relevant for Parkinson’s disease detection are inevitably discarded during the cepstral transformation process. This may explain the lower performance observed when MFCCs constitute the sole handcrafted acoustic representation available to the model.

The complete three-branch architecture consistently achieves the best results across all evaluation metrics, obtaining an accuracy of 91.51\%, an F1-score of 91.24\%, and a recall of 89.00\%. Compared with the strongest two-branch configuration (Spectrogram + HuBERT), the proposed methodology improves accuracy by more than four percentage points and increases the F1-score by approximately 4.6 percentage points. Furthermore, the complete model exhibits lower standard deviations across folds, indicating improved robustness and stability under speaker-independent evaluation conditions.

These findings demonstrate that the spectrogram and MFCC branches provide complementary information that is not fully captured by the HuBERT embeddings alone. While spectrograms contribute detailed time-frequency information and MFCCs encode vocal tract characteristics, HuBERT provides high-level contextual representations learned through self-supervised pre-training. The proposed context-guided cross-modal attention framework is able to effectively exploit the synergy between these heterogeneous representations, allowing the model to capture pathological speech characteristics at multiple levels of abstraction. Consequently, the best performance is achieved when all three modalities are jointly incorporated into the diagnostic framework.

\begin{table}[h!]
\caption{ABLATION STUDY (MODALITIES). BEST RESULTS PER EVALUATION METRIC ARE IN BOLD.}
\label{ablation_modalities}
\begin{center}
\resizebox{\columnwidth}{!}{%
\begin{tabular}{lcccccc}
\toprule
& \multicolumn{6}{c}{\textbf{Evaluation metrics}} \\
\cline{2-7}
\textbf{Architecture} & \textbf{Precision} & \textbf{Recall} & \textbf{F1-score} & \textbf{Accuracy} & \textbf{Specificity} & \textbf{AUC} \\
\midrule
\rowcolor{gray!30} \multicolumn{7}{l}{\textbf{Ablation Experiments}} \\
Spectrogram + HuBERT Base & 92.43 & 82.00 & 86.62 & 87.45 & 92.95 & 95.17 \\
                          & $\pm$5.14 & $\pm$8.12 & $\pm$5.08 & $\pm$4.62 & $\pm$5.08 & $\pm$2.05 \\
\hline
MFCC + HuBERT Base        & 90.36 & 81.00 & 85.16 & 85.98 & 90.99 & 94.92 \\
                          & $\pm$4.45 & $\pm$6.63 & $\pm$3.57 & $\pm$3.04 & $\pm$4.90 & $\pm$1.91 \\
\midrule
\rowcolor{gray!30} \multicolumn{7}{l}{\textbf{Proposed Methodology}} \\
\textbf{Methodology}      & \textbf{93.99} & \textbf{89.00} & \textbf{91.24} & \textbf{91.51} & \textbf{93.89} & \textbf{95.97} \\
                          & $\pm$3.62 & $\pm$4.90 & $\pm$1.51 & $\pm$1.14 & $\pm$3.86 & $\pm$2.32 \\
\bottomrule
\end{tabular}%
}
\end{center}
\end{table}

\section{Conclusion}

In this article, we introduce a novel multi-branch deep learning framework for the automated detection of PD from speech. Unlike conventional approaches that rely on a single acoustic representation, the proposed architecture jointly exploits three complementary views of the speech signal, namely Log-Mel spectrograms, MFCC sequences, and HuBERT Base embeddings. These heterogeneous representations are integrated through a context-guided cross-modal attention mechanism, which leverages spectral and cepstral information to identify the most diagnostically relevant temporal regions within the self-supervised HuBERT representations. Experimental results on the PC-GITA corpus demonstrate the effectiveness of the proposed approach, achieving a subject-level accuracy of 91.51\%, an F1-score of 91.24\%, and an AUROC of 95.97\% under a speaker-independent evaluation protocol. Furthermore, the ablation study confirms the contribution of each modality and validates the effectiveness of the proposed multimodal fusion strategy.

Future work will focus on evaluating the cross-lingual and cross-dataset generalization capabilities of the proposed architecture on additional multilingual Parkinson’s disease speech corpora. We also plan to investigate federated learning paradigms, where individual datasets are treated as separate clients, enabling collaborative model training while preserving data privacy and facilitating the development of more robust and generalizable PD detection systems.
\bibliographystyle{IEEEtran}
\bibliography{references}

\end{document}